# Epitaxial Growth of the Diluted Magnetic Semiconductors $Cr_yGe_{1-y}$ and $Cr_yMn_xGe_{1-x-y}$

G. Kioseoglou,[a] A.T. Hanbicki, C.H. Li,[b] S.C. Erwin, R. Goswami,[c] and B.T. Jonker
*Naval Research Laboratory*, Washington, DC 20375

(Submitted 1/13/03)

We report the epitaxial growth of $Cr_yGe_{1-y}$ and $Cr_yMn_xGe_{1-x-y}$(001) thin films on GaAs(001), describe the structural and transport properties, and compare the measured magnetic properties with those predicted by theory. The samples are strongly p-type, and hole densities increase with Cr concentration. The $Cr_yGe_{1-y}$ system remains paramagnetic for the growth conditions and low Cr concentrations employed (y ≤ 0.04), consistent with density functional theory predictions. Addition of Cr into the ferromagnetic semiconductor $Mn_xGe_{1-x}$ host systematically reduces the Curie temperature and total magnetization.

Ferromagnetic semiconductors (FMS) are promising materials for semiconductor- based spintronic device applications. They can be easily incorporated in semiconductor heterostructures, and offer new functionality and avenues for spin manipulation [1]. The coexistence of semiconductor properties and spontaneous long-range ferromagnetic (FM) order observed in these systems provides opportunity for fundamental studies as well as a variety of magnetic and spin-dependent devices. Although much effort has focused on III-Mn-V [1-4] materials, the mechanism of FM order remains unclear, particularly the precise role played by dopants and the semiconductor host. FM order has also been predicted to occur in other hosts [5,6]. Ge provides a simple host lattice to explore the fundamental origins of FM order, and it is lattice matched to the AlGaAs/GaAs family, facilitating incorporation into III-V heterostructures. Recently, the epitaxial growth of a group IV FMS, $Mn_xGe_{1-x}$, has been reported with Curie temperatures ($T_c$) up to 116 K [7]. Voltage controlled FM order in $Mn_xGe_{1-x}$ has been demonstrated in simple gated structures using very low gate voltages (±0.5 V) [7].

This recent work motivates efforts to investigate other Ge-based FMS candidates and test predictive models. We report here the epitaxial growth of $Cr_yGe_{1-y}$ and $Cr_yMn_xGe_{1-x-y}$ thin films, and compare the measured magnetic properties with those predicted by theory. Like Mn, Cr is expected to produce p-type material with acceptor levels at 70 meV and 120 meV above the Ge valence band edge. A large hole density is a necessary criterion for stabilizing FM order in the mean-field Zener model [5,6]. Our density-functional theory (DFT) calculations indicate that the substitutional site is favored for both Cr and Mn impurities in the Ge host lattice – the formation energies for the substitutional and tetrahedral interstitial sites are 2.1 and 2.4 eV for Cr, and 1.2 and 1.9 eV for Mn, respectively. The Cr-Ge bulk phase diagram for low Cr concentrations [8] shows two stable phases; a Ge with negligible amount of Cr, and an intermetalic compound, $Cr_{11}Ge_{19}$, the only compound that exhibits FM order with a Curie temperature of 91 K [9].

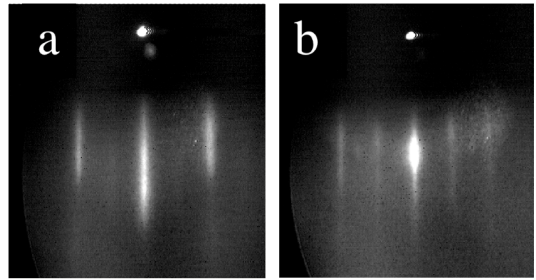

Fig. 1. RHEED patterns for two $Cr_yGe_{1-y}$ samples: (a) $T_s$=200 °C, y=0.022; (b) $T_s$=400 °C, y=0.033. The electron beam is incident along the [100] and [110] azimuths, respectively.

$Cr_yGe_{1-y}$ samples were grown on semi-insulating (SI) GaAs (001) substrates using molecular beam epitaxy (MBE) at a growth rate of ~ 5 Å / min. Sample thickness and composition were determined from x-ray fluorescence measurements. A series of $Cr_yGe_{1-y}$ samples was grown at substrate temperatures ($T_s$) in the range of 40 to 500 °C with 0.01<y<0.04. Reflection high-energy electron diffraction (RHEED) was used to monitor the quality of growth, and patterns for two of the samples (one grown at $T_s$= 200 °C and the other at $T_s$ =400 °C) are shown in Fig.1. The patterns exhibit a streaky 2x1 reconstruction throughout the growth for $T_s$ > 200 °C, indicating predominantly 2-dimensional growth. Growth at lower substrate temperatures was initially single crystal, but the pattern became increasingly diffuse with film thickness..

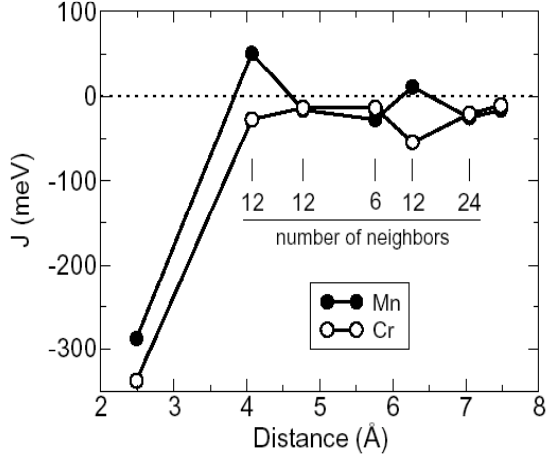

Fig. 2. Calculated spin coupling constants from density-functional theory for Cr-Cr and Mn-Mn interactions in a Ge host as a function of separation. FM (AFM) coupling corresponds to positive (negative) values.

Hall measurements, using the van der Pauw geometry at 300 K, reveal high hole concentrations ($10^{19}$-$10^{21}$ cm$^{-3}$), which increase with Cr concentration. These hole densities are significantly higher than those observed in the $Mn_xGe_{1-x}$ system [7], consistent with the shallower acceptor level for Cr (70 meV vs. 160 meV for Mn). Magnetization measurements using SQUID magnetometry show that all $Cr_yGe_{1-y}$ samples are paramagnetic in the temperature range of 4-300 K. There is no evidence for the formation of the known ferromagnetic phase $Cr_{11}Ge_{19}$ ($T_c$=91K) for the growth conditions employed here.

These results are in good agreement with predictions based on DFT calculations. We computed the spin coupling between Cr dopants in a Ge host, using 128-atom supercells with two substitutional Cr dopants. Cr-Cr separations from nearest-neighbor up to ~8 Å were considered, and ideal diamond-lattice positions were used for all calculations. Spin-spin coupling constants, J(R), were computed from the total-energy differences between parallel and antiparallel spin arrangements. The calculations were performed within the generalized-gradient approximation [10] to DFT, using ultrasoft pseudopotentials in a plane-wave basis with a cutoff energy of 227 eV [11,12]. The Brillouin zone-center was sufficient to converge total-energy differences to a few meV.

The resulting Cr-Cr spin coupling constants J(R) are shown in Fig.2 (open circles). For comparison, we also show the results of identical calculations for Mn-Mn interactions in Ge (Fig.2 solid circles). Both dopants have a strong antiferromagnetic (AFM) interaction at nearest-neighbor separation. For larger separations, the interactions are much smaller in magnitude and show a large crystallographic anisotropy that most likely originates from the $l$=2 angular momentum parentage of the $t_2$ impurity wavefunction. For Mn, the interaction is FM at the second and fifth neighbor shells; this is apparently sufficient to stabilize the long-range ferromagnetic order found experimentally up to ~ 120 K in the doped crystals [7]. In contrast, the interactions for Cr are AFM for all separations considered. Calculations by Kudrnovsky and co-workers show similar behavior [13]. Hence, barring any qualitative change in J(R) with increasing Cr concentration (i.e. a sign change), and assuming there are no significant ferromagnetic interactions beyond 8 Å, our calculations predict that uniformly Cr-doped Ge should exhibit no ferromagnetic phase at any temperature, as found here experimentally..

Choi et al recently reported FM order in bulk grown $Cr_{0.01}Ge_{0.99}$ with a nominal $T_c$ = 126 K [14], in contrast with both our experimental and theoretical results described above. However, a comparison of their data with standard reference data [9] strongly suggests the presence of the known bulk FM phase $Cr_{11}Ge_{19}$ ($T_c$=91 K) in their material. The higher value of $T_c$ they report, which is the basis of their argument for ruling out the presence of $Cr_{11}Ge_{19}$, appears to be a consequence of how they choose to define $T_c$.

Since Cr provides exceptionally high hole densities in Ge, we attempted to use it as a p-dopant in the ferromagnetic $Mn_xGe_{1-x}$ host with the intent of enhancing the hole density and the corresponding Curie temperature. The Mn concentration was fixed at 2.5% while the nominal Cr concentration was varied from 1 to 4%. The $Cr_yMn_xGe_{1-x-y}$ samples were grown on SI GaAs (001) substrates to a typical thickness of 1000 Å. After growing a Ge buffer layer at 350 °C, the substrate temperature was reduced to 70 °C for $Cr_yMn_xGe_{1-x-y}$ growth. This procedure was used to minimize phase separation or the formation of unwanted compounds in $Mn_xGe_{1-x}$ [7]. RHEED patterns exhibit a 2x1 reconstruction throughout the growth of both the buffer and the $Cr_yMn_xGe_{1-x-y}$ film, indicating good crystallinity.

Magnetization loops show hysteretic behavior and substantial remanence with saturation magnetizations up to 10 emu/cm$^3$. No detectable FM signal is observed near room temperature, confirming the absence of any known FM phases of Mn:Ge such as $Mn_{11}Ge_8$ [15]. In the inset of Fig.3 we plot the magnetic field dependence of the magnetization for sample 515B (3.3% nominal Cr concentration) at T=5K, with the field applied in-plane. Clear hysteresis with a coercive field of 1000 Oe is observed despite the large content of Cr. The saturation magnetization for this sample, $M_s$=7.2 emu/cm$^3$ (=0.009T), corresponds to 2.5 x 10$^{20}$/cm$^3$ active magnetic ions (assuming S=3/2 for Mn) compared to a total Mn concentration of 11 x 10$^{20}$/cm$^3$ (2.5%). Figure 3 summarizes the temperature dependence of the magnetization for a series of samples with increasing Cr concentration. The addition of Cr systematically reduces both $T_c$ and the saturation magnetization. Since the magnetization curves (Fig.3) differ from the classic-ferromagnetic Brillouin behavior, $T_c$ was estimated from a Curie-Weiss fit to the high temperature data. The results are summarized in Fig.4. $T_c$ is monotonically reduced with increasing Cr concentration from 120 K for the reference Mn:Ge sample to 40 K for the sample with 4.1% Cr.

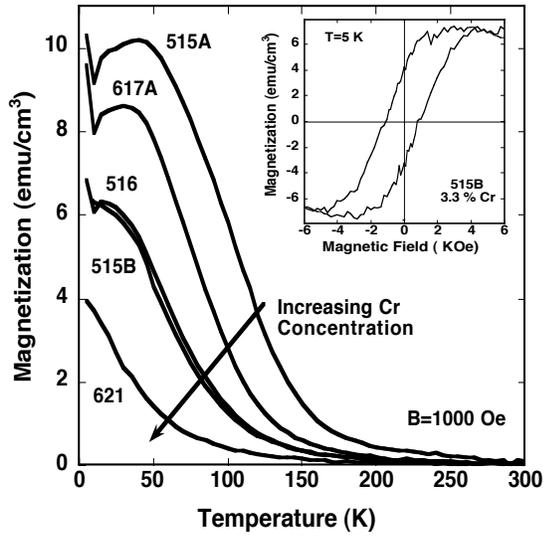

Fig. 3. Temperature dependence of the magnetization for the $Cr_yMn_xGe_{1-x-y}$ samples. The inset shows the B-H loop for sample 515B.

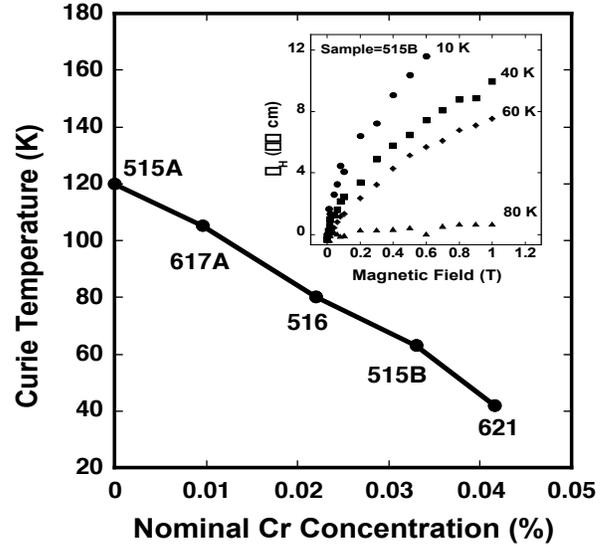

Fig. 4. Dependence of Curie Temperature on nominal Cr concentration. The inset shows a plot of Hall resistivity for sample 515B at different temperatures- a pronounced EHE is observed below $T_c$.

A pronounced extraordinary Hall effect (EHE), another indication of ferromagnetism, persists despite suppression of FM order with increasing Cr concentration. The inset of Figure 4 shows the Hall resistivity plotted as function of the applied magnetic field for several temperatures below and above $T_c$ for sample 515B ($T_c$ = 63 K). The EHE dominates at low temperatures and low fields, but is essentially absent for temperatures higher than $T_c$. Although the EHE makes a determination of hole densities less accurate, an estimate can be obtained from the Hall data at high magnetic fields where the magnetization is saturated. As an example, for sample 515B (Fig.4 inset) we obtain p = 4 x 10$^{20}$/cm$^3$ at T=40 K, which is comparable to the effective Mn ion concentration of 2.5x10$^{20}$/cm$^3$ deduced from the total magnetization as discussed in the preceding paragraph. Note that this is much larger hole density than 1.7 x 10$^{19}$/cm$^3$ obtained from EHE data for the reference Mn:Ge sample (515A). Thus Cr successfully serves as a co-dopant and enhances the hole density, but nevertheless suppresses FM order in the system.

In an effort to better understand the role played by Cr in the $Cr_yMn_xGe_{1-x-y}$ system we performed microstructural analysis using transmission electron microscopy (TEM). The TEM images and diffraction data reveal the formation of a second phase in the Ge matrix, $Ge_3Cr_5$, a paramagnetic compound. The formation of such a paramagnetic phase is consistent with the trends observed in the magnetization data. A detailed discussion of these data and a structural analysis will be presented elsewhere [16].

In summary, we have grown $Cr_yGe_{1-y}$ and $Cr_yMn_xGe_{1-x-y}$ single crystal films to further investigate the Ge-based diluted magnetic semiconductor family. The Cr:Ge system is paramagnetic for the growth conditions and low Cr concentrations employed, and this result is consistent with the behavior predicted by DFT calculations. The addition of Cr in the ferromagnetic semiconductor $Mn_xGe_{1-x}$ host systematically reduces the Curie temperature and the total magnetization.

This work was supported by the Office of Naval Research (L. Cooper) and the DARPA *Spins in Semiconductors* program (S. Wolf).

[1] H. Ohno, Science **281**, 951 (1998); S.A. Wolf, D.D. Awschalom, R.A. Buhrman, J.M. Daughton, S. von Molnar, M.L. Roukes, A.Y. Chtchelkanova, and D.M. Treger, Science **294**, 1488 (2001).

[2] H. Munekata, H. Ohno, S. von Molnar, Armin Segmuller, L.L. Chang, and L. Esaki, Phys. Rev. Lett. **63**, 1849 (1989).

[3] H. Ohno, H. Munekata, T. Penny, S. von Molnar, and L.L. Chang, Phys. Rev. Lett. **68**, 2664 (1992).

[4] H. Ohno, A. Shen, F. Matsukura, A. Oiwa, A. Endo, S. Katsumoto, and Y. Iye, Appl. Phys. Lett. **69**, 363 (1996).

[5] T. Dietl, H. Ohno, F. Matsukura, J. Cibert, and D. Ferrand, Science **287**, 1019 (2000).

[6] T. Dietl, H. Ohno, and F. Matsukura, Phys. Rev. B**63**, 195205 (2001).

[7] Y.D. Park, A.T. Hanbicki, S.C. Erwin, C.S. Hellberg, J.M. Sullivan, J.E. Mattson, T. Ambrose, A. Wilson, G. Spanos, and B.T. Jonker, Science **295**, 651 (2002).

[8] A.B. Gokhale and G.J. Abbaschian, Bulletin of Alloy Phase Diagram, **7**, 477 (1986).

[9] LANDOLT-BORNSTEIN, *Numerical Data and Functional Relationships in Science and Technology,* New Series III/19c, edited by O. Madelung, (Springer-Verlag, Berlin, 1988) p.12.

[10] J.P. Perdew and Y. Wang, Phys. Rev. B**45**, 13244 (1992).

[11] G. Kresse and J. Hafner, Phys. Rev. B**47**, 558 (1993).

[12] G. Kresse and J. Furthmueller, Phys. Rev. B**54**, 11169 (1996).

[13] J. Kudrnovsky, *2$^{nd}$ International Conference on Physics and Application of Spin Related Phenomena in Semiconductors*, July 2002, Wurzburg, Germany.

[14] S. Choi, S.C. Hong, S. Cho, Y. Kim, J.B. Ketterson, C.U. Jung, K. Rhie, B.J. Kim, and Y.C. Kim, Appl. Phys. Lett. **81**, 3606 (2002).

[15] Y.D. Park, A. Wilson, A.T. Hanbicki, J.E. Mattson, T. Ambrose, G. Spanos, and B.T. Jonker, Appl. Phys. Lett. **78**, 2739 (2001).

[16] R. Goswami, G. Kioseoglou, A.T. Hanbicki, A. Wilson, B.T. Jonker, and G. Spanos, [to be published].